\documentclass[prx,superscriptaddress,nofootinbib]{revtex4}
\usepackage[latin9]{inputenc}
\usepackage{color}
\usepackage{amsmath}
\usepackage{amssymb}
\usepackage{graphicx}
\usepackage{tikz}
\usepackage{verbatim}
\usepackage{bm}

\usepackage{amsmath}
\usepackage{amssymb}
\usepackage{graphicx}
\usepackage{tikz}
\usepackage{amsmath,bm,amssymb,latexsym,galois,amsthm,euscript,dsfont}
\usepackage[all,cmtip,knot]{xy}
\usepackage{dsfont} 
\usepackage{mathrsfs}
\usepackage[unicode=true,bookmarks=true,bookmarksnumbered=false,bookmarksopen=false,breaklinks=false,pdfborder={0 0 1},backref=false,colorlinks=true]{hyperref}

\hypersetup{
    colorlinks,
    linkcolor={red!50!black},
    citecolor={blue!90!black},
    urlcolor={blue!90!black}
}

\makeatletter
\@ifundefined{textcolor}{}
{%
 \definecolor{BLACK}{gray}{0}
 \definecolor{WHITE}{gray}{1}
 \definecolor{RED}{rgb}{1,0,0}
 \definecolor{GREEN}{rgb}{0,1,0}
 \definecolor{BLUE}{rgb}{0,0,1}
 \definecolor{CYAN}{cmyk}{1,0,0,0}
 \definecolor{MAGENTA}{cmyk}{0,1,0,0}
 \definecolor{YELLOW}{cmyk}{0,0,1,0}
}

\usepackage{amsfonts}\usepackage{tabularx}\usepackage{dcolumn}\usepackage{bm}\usepackage{graphicx}\usepackage{epstopdf}

\hypersetup{urlcolor=blue}

\theoremstyle{definition}
\newtheorem{defn}{Definition}

\newtheorem{exmp}[defn]{Example}

\newtheorem*{defn*}{Definition}

\theoremstyle{remark}

\makeatother

\begin{document}

\title{Quantum advantages of communication complexity from Bell nonlocality}

\author{Zhih-Ahn Jia}
\email{giannjia@foxmail.com}
\affiliation{CAS Key Laboratory of Quantum Information, School of Physical Sciences, University of Science and Technology of China, Hefei, Anhui, 230026, P.R. China}
\affiliation{CAS Center For Excellence in Quantum Information and Quantum Physics, University of Science and Technology of China, Hefei, Anhui, 230026, P.R. China}

\author{Lu Wei}
\email{luweiphys@gmail.com}
\affiliation{School of the Gifted Young, University of Science and Technology of China, Hefei, Anhui, 230026, P.R. China}

\author{Yu-Chun Wu}
\email{wuyuchun@ustc.edu.cn}
\affiliation{CAS Key Laboratory of Quantum Information, School of Physical Sciences, University of Science and Technology of China, Hefei, Anhui, 230026, P.R. China}
\affiliation{CAS Center For Excellence in Quantum Information and Quantum Physics, University of Science and Technology of China, Hefei, Anhui, 230026, P.R. China}

\author{Guang-Can Guo}
\email{gcguo@ustc.edu.cn}
\affiliation{CAS Key Laboratory of Quantum Information, School of Physical Sciences, University of Science and Technology of China, Hefei, Anhui, 230026, P.R. China}
\affiliation{CAS Center For Excellence in Quantum Information and Quantum Physics, University of Science and Technology of China, Hefei, Anhui, 230026, P.R. China}

\begin{abstract}
Communication games are crucial tools for investigating the limitations of physical theories. 
The communication complexity (CC) problem is a typical example, for which several distributed parties attempt to jointly calculate a given function with limited classical communications. 
In this work, we present a method to construct CC problems from Bell tests in a graph-theoretic way.
Starting from an experimental compatibility graph and the corresponding Bell test function, a target function which encodes the information of each edge can be constructed, then using this target function we could construct an CC function for which by pre-sharing entangled states, the success probability will exceed that for arbitrary classical strategy.  
The non-signaling protocol based on Popescu-Rohrlich box is also discussed, and the success probability in this case would reach one. 
\end{abstract}
\maketitle

\section{Introduction}
\label{sec:introduction}

The Bell nonlocality \cite{einstein1935,bell1964,Horodecki2009,Brunner2014bell} is one of the most distinctive features that distinguish quantum mechanics from classical mechanics.
It's an experimentally verified phenomenon and now serves as a crucial resource for many quantum information tasks,  such as quantum computation \cite{Nielsen2010}, quantum key distribution (QKD) \cite{Ekert1991quantum}, quantum random number generator \cite{HerreroCollantes2017quantum}, communication complexity (CC) problem \cite{Buhrman2010} and so on.
Among these tasks, CC problems  for which distributed parties jointly calculate a function with limited communications are of great importance for investigating the limitations of different physical theories  \cite{Buhrman2010,Brunner2014bell}. For instance, the set of calculable functions and the success probabilities for calculating a given function may be different for local hidden variable (LHV) theory \cite{bell1964}, quantum theory and non-signaling theory.


The CC problems, originally introduced by Yao \cite{Yao1979some}, concern the question what is the minimal amount of communication necessary for two or more parties to jointly calculate a given multivariate function $f(x_1,\cdots,x_n)$ where the $k$-th party only knows his own input $x_k$ but no information about the inputs of other parties initially. It has been shown from different perspectives that entanglement and Bell nonlocality are closely related to the quantum advantage of CC problem, see Refs. \cite{brukner2004bell, degorre2009communication, junge2018classical, tavakoli2020does, Buhrman2010, buhrman2016quantum, laplante2018robust}. 
Violation of Bell inequalities often leads to quantum advantages of the CC problems \cite{brukner2004bell, Buhrman2010,degorre2009communication,junge2018classical, tavakoli2020does} and it's also argued that quantum advantage of CC problem implies the violation of Bell inequalities \cite{buhrman2016quantum,laplante2018robust}. 
However, many of the results above are existence proof. In practice, to utilize Bell nonlocality to obtain quantum advantages of a real CC problem, one needs to consider how to explicitly translate the Bell test into a CC problem. In this work, we systematically explore the translation of a Bell test into CC problem via graph-theoretic method.

We are mainly concerned with the CC problems for which only limited  classical communications are allowed, and the goal for each party is to calculate the function with as high success probability as possible. 
By introducing the concept of the experimental compatibility graph and its corresponding Bell test function, we explore the relationship between the Bell nonlocality and quantum advantage of CC problems. 
We show that, from an arbitrary experimental compatibility graph $G^e$, we can construct a corresponding CC problem $F_{G^e}$, for which the quantum protocol exhibits a success probability that exceeds the success probabilities for all classical protocols. 
We also investigate the possibility of using non-signaling box for solving CC problems, and we show that it has an advantage over all quantum protocols.

The paper is organized as follows, in Sec. \ref{sec:bell}, we introduce several graph-theoretic concepts related to Bell nonlocality, including the experimental compatibility graph, compatibility graph, and the Bell test functions. 
In Sec. \ref{sec:QCC}, we give the basics of CC problems and define the quantum advantages of the protocol. 
In Sec. \ref{sec:qadv}, we present a class of functions based on arbitrary given experimental  graph $G^e$, for which quantum protocols exhibit advantages. Finally, in the last section, we give some concluding remarks.

\section{Bell inequalities from compatibility graphs}
\label{sec:bell}
\begin{figure}
  \centering
    \includegraphics[width=10cm]{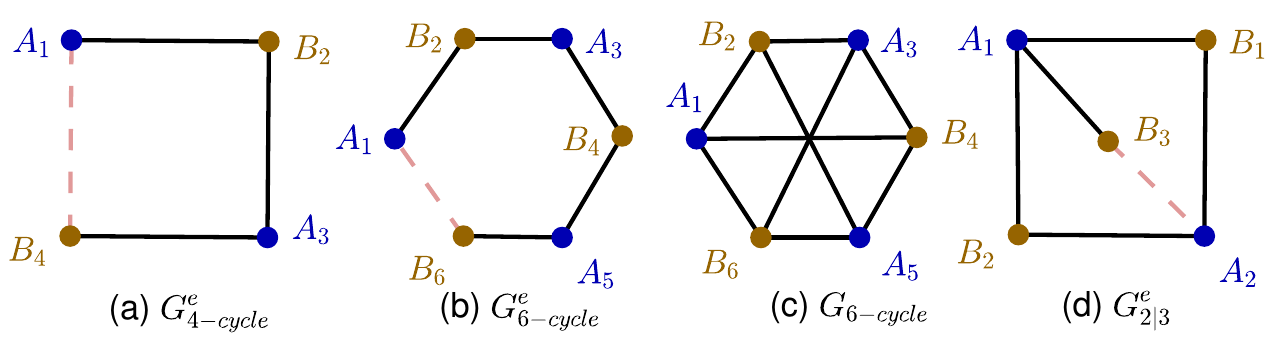}\\
  \caption{The depiction of the experimental  compatibility graph and measurement compatibility graph. (a) the experimental compatibility graph $G^e_{4-cycle}$ of CHSH inequality, it is a 4-cycle; (b) the experimental compatibility graph $G^e_{6-cycle}$ of 6-cycle Bell inequality, it is a 6-cycle; (c) The  measurement compatibility graph $G_{6}$ corresponds to  $G^e_{6-cycle}$; (d) a non-cycle experimental compatibility graph $G^e_{2|3}$.}\label{fig:cycle}
\end{figure}

Let us now introduce a general framework for $n$-party Bell inequalities based on a set of $n$-point correlation functions $E(x_{i_1},x_{i_2},\cdots,x_{i_n})=\langle x_{i_1}\otimes x_{i_2}\otimes \cdots \otimes x_{i_n} \rangle$. 
Many pertinent classes of Bell inequalities are of this correlator form, see, e.g., Refs.  \cite{Brunner2014bell,CHSH}.
To start with, let us first introduce a useful mathematical tool, compatibility graphs. 
For a set of measurements $\mathcal{M}=\{M_1,\cdots,M_n\}$, we can assign a corresponding graph $G_{\mathcal{M}}$, called \emph{measurement compatibility  graph} \cite{Jia2016}, whose vertices are labeled by measurements and there is an edge between two vertices if the corresponding measurements are compatible, i.e., they can be measured simultaneously. We denote the vertex set of the graph $G$ as $V(G)$ and the edge set as $E(G)$, and an edge is a pair $\langle ij\rangle :=(M_i,M_j)\in V(G)\times V(G)$.
Similarly, we can introduce the \emph{experimental compatibility graph and hypergraph} $G^e_{\mathcal{M}}$\cite{Jia2016}, in which the vertices
are labeled with the measurements involved in the experiment, and an edge represents two or more jointly measured measurements in the experiment.
For two-party case, each edge consists of two measurements,  $G^e_{\mathcal{M}}$ is a subgraph of $G_{\mathcal{M}}$; for $n$-party ($n>2$) case, each edge consists of $n$ vertices, thus $G^e_{\mathcal{M}}$ is a hypergraph.
See Fig. \ref{fig:cycle} (b) and (c) for illustration of compatibility graph and two-party experimental compatibility graph.

In a typical $n$-party Bell scenario,  the experimenters share an $n$-partite system. According to an experimental compatibility  graph $G^e$, they can choose a set of measurements $x_{i_1},x_{i_2},\cdots,x_{i_n}$ to jointly measure, where $x_{i_k}$ is the measurement chosen by the $k$-th party.
After many runs of experiments, they obtain a set of $n$-point correlation functions $\{E(x_{i_1},x_{i_2},\cdots,x_{i_n})| \langle i_1i_2\cdots i_n\rangle \in E(G^e)\}$.
To test if the obtained measurement statistics are local, viz., obey the LHV theory or not, they need to calculate a function 
\begin{equation}
	\begin{aligned}
\mathcal{B}_{G^e}=&\sum_{\langle i_1i_2\cdots i_n\rangle\in E(G^e)}\gamma_{\langle i_1i_2\cdots i_n \rangle}E(x_{i_1},x_{i_2},\cdots,x_{i_n})     \\
&+\sum_{\langle i_1i_2\cdots i_{n-1}\rangle\in E(G^e)}\gamma_{\langle i_1i_2\cdots i_{n-1} \rangle}E(x_{i_1},x_{i_2},\cdots,x_{i_{n-1}})\\
&+\cdots +\sum_{x_i\in V(G) } \gamma_{\langle x_i \rangle} E(x_i),
\end{aligned}
\end{equation}
which we refer to as \emph{Bell test function}. 
In this work, we will mainly focus on the homogenous case, namely, for $n$-party Bell test, the Bell test function only contain $n$-point correlation functions.
It's also convenient for our purpose to assume that $\gamma_{\langle  i_1i_2\cdots i_n\rangle}=\pm 1$. In this case, different colors of edges of $G^e$ represents different coefficients, if $\gamma_{\langle  i_1i_2\cdots i_n\rangle}=+1$ the edge is drawn as black solid line and called positive edge, and if $\gamma_{\langle  i_1i_2\cdots i_n\rangle}=-1$, the edge is drawn as red dashed line and called negative edge, as depicted in Fig. \ref{fig:cycle}.

Note that for an $n$-party Bell experiment, $G^e$ is usually an $n$-partite graph, since the measurements of each party are usually chosen as incompatible measurements. 
In a LHV world, the value of the test function lies in a range $\mathcal{R}_{LHV}=[B_C^1,B_C^2]$, e.g., for Clauser-Horne-Shimony-Holt (CHSH) type Bell test function $\mathcal{B}_{CHSH}$, the range is $\mathcal{R}_{LHV}=[-2,2]$ \cite{CHSH}. But for quantum theory, the value may lie outside the LHV range  $\mathcal{R}_{LHV}$, which is called the quantum violation of the Bell inequality, which means that quantum theory is not consistent with the LHV assumption. Similar to LHV theory, there also exists a quantum range $\mathcal{R}_Q=[B_Q^1,B_Q^2]$ of the value of Bell test function, e.g., for CHSH type Bell test function, it is  $\mathcal{R}_{Q}=[-2\sqrt{2},2\sqrt{2}]$, and this kind of quantum bound is known as Tsirelson bound. If it is possible for a Bell test function to violate the quantum range? 
The answer is yes, there are many different kinds of approaches to understand quantum theory  from outside, e.g., in non-signaling theory \cite{Popescu1994}, the Bell test function may reach its functional minimal and maximal values. To summarize, we have the Bell inequalities for a given experiment compatibility graph as
\begin{equation}
\mathcal{B}_{G^e}\overset{LHV}{\in}\mathcal{R}_{LHV}\overset{Q}{\subseteq} \mathcal{R}_{Q}\overset{NS}{\subseteq} \mathcal{R}_{NS}.
\end{equation}
Note that for a given experimental compatibility graph, the Bell test function is in general not unique.

Another crucial issue is what kind of  experimental compatibility graph can be used to test Bell nonlocality.  A necessary condition for this  is the following \cite{Ramanathan2012generalized,Jia2016}: the compatibility graph corresponding to $G^e$ is non-chordal. Chordal graphs are those that do not have any induced cycle of size more than three. 
From Vorob'yev theorem \cite{vorob1963markov,vorob1967coalitional}, 
if the compatibility graph $G$ corresponding to $G^e$ is chordal, then there always exists a global joint probability distribution which can reproduces all marginal probability distributions we obtained from the experiment. A result of Fine \cite{Fine1982} further claims that the existence of this kind of global joint probability distribution 
 is equivalent to the existence of a LHV model for all involved measurements. 
 Thus for chordal graph, the measurement statistics are always reproducible by the LHV model. 
In a recent work \cite{Cabello2019necessary}, it's claimed that the above condition is also a sufficient condition.

Here we present two examples for convenience of our later discussions. We recommend readers to read Refs. \cite{Ramanathan2012generalized,Jia2016,jia2017exclusivity,jia2018,Cabello2019necessary} for more examples.

\begin{exmp}\label{exp:1}
The first example is $2m$-cycle Bell inequality. The experimental compatibility graph is a $2m$-cycle, for which $A_1, A_3, \cdots, A_{2m-1}$ are observables chosen by Alice and $B_2,B_4,\cdots,B_{2m}$ are observables chosen by Bob, the $i$-th vertex connects with the $(i+1)$-th vertex. 
The Bell test function is thus
\begin{align}
\mathcal{B}_{G^e_{\mathrm{cycle}}}=&\sum_{i=1,3,\cdots,2m-1}\gamma_{i}E(A_i,B_{i+1})+\sum_{i=2,4,\cdots,2m}\gamma_{i}E(A_{i+1},B_{i})
\end{align}
note that here $\gamma_{i}=\pm 1$, the the number of $\gamma_i=-1$ must be odd to ensure that it can test Bell nonlocality.
As proved in  \cite{braunstein1988information,Wehner2006tsirelson},  the Bell inequality is
\begin{equation}
|\mathcal{B}_{G^e_{\mathrm{cycle}}}|\overset{LHV}{\leq} 2m-2\overset{Q}{\leq} 2m \cos \frac{\pi}{2m}\overset{NS}{\leq} 2m.
\end{equation}
When $m=2$, the experimental  compatibility graph is a 4-cycle as depicted in Fig. \ref{fig:cycle} (a), the corresponding Bell inequality is the CHSH inequality.
\end{exmp}
 It's worth mentioning that, although we can construct a Bell test from arbitrary non-chordal graph, the LHV bound (which corresponds to independent number calculation of a graph) and non-signaling boundary are easy to obtain, but the maximum quantum violation (which corresponds to the Lov\'{a}sz number calculation of a graph\cite{Cabello2014graph,jia2017exclusivity}) is usually very difficult to calculate.
The example corresponding to non-cycle experimental compatibility graph can also be constructed. 
\begin{exmp}\label{exp:2}
The experimental compatibility  graph of this Bell test is shown in Fig. \ref{fig:cycle} (d). We denote the graph as $G^e_{2|3}$, and the subscripts here is used to indicate that Alice chooses two observables and Bob chooses three observables to  measure. 
The corresponding Bell test function is 
\begin{align}
\mathcal{B}_{G_{2|3}^e} = &E(A_1,B_1)+E(A_2,B_1)+E(A_2,B_2)+E(A_1,B_2) \nonumber\\
&+E(A_1,B_3)-E(A_2,B_3).
\end{align}
The LHV bound is $4$ and the non-signaling bound is $6$, but the exact quantum bound is still unknown.
\end{exmp}

\section{Communication complexity problems}
\label{sec:QCC}

Now, let's  recall the formal definition of communication complexity, for more details, we refer the reader to Refs. \cite{kushilevitz_nisan_1996,arora2009computational,hromkovivc2013communication}. For simplicity, consider two-party case, for which Alice and Bob try to calculate a bivariate function $f:\mathbb{B}^n\times \mathbb{B}^n\to \mathbb{B}$ collaboratively, where $\mathbb{B}$ denotes binary set $\{0,1\}$ or $\{\pm 1\}$. An $r$-round communication complexity protocol $\mathcal{P}$ for computing function $f(x,y)$ is a distributed algorithm consisting of a set of $r$ functions $f_1,\cdots,f_r:\cup_{m\geq 0} \mathbb{B}^m \to \cup_{m\geq 0}  \mathbb{B}^m$. 
Alice first individually calculates function $f_1(x)=v_1$ and sends the result to Bob, after Bob receives the result, he calculates function $f_2(y,v_1)=v_2$ and sends the result to Alice, etc. 
Each communication between them is called a round. 
We say the protocol $\mathcal{P}$ is valid for calculating $f(x,y)$ if the last message sent (i.e., $v_r=f_r(x,v_1,\cdots,v_{r-1})$ by Alice or $v_r=f_r(y,v_1,\cdots,v_{r-1})$ by Bob) equals to $f(x,y)$ for all possible input values of $x,y$. The communication complexity of the protocol $\mathcal{P}$ is  then defined as the $C_{\mathcal{P}}(f)=|v_1|+\cdots +|v_r|$, where $|v_i|$ denotes the number of bits of the message $v_i$. The protocol defined above is deterministic. For bounded-error case, Alice and Bob can toss coins individually or jointly to choose the input at each round, and the protocol $\mathcal{P}$ has to calculate $f$ with success probability greater than or equal to a fixed value $1-\delta$ where $\delta$ is usually chosen as $1/3$, viz, $P_{succ}\geq 2/3$. We assume that Bob guesses the value $f(x,y)$ as $z=0,1$ at the final round, the successful probability will be
\begin{equation}
p_{succ}(\mathcal{P})=\sum_{x,y}p(x,y)p(z=f(x,y)|x,y).
\end{equation}
The bounded-error communication complexity is denoted as $C^{be}_{\mathcal{P}}(f)$ which is the number of communicated bits in the protocol such that $p_{succ}\geq 1-\delta$ for some $\delta< 1/2$.

Bounded-error communication complexity problem concerns the problem of getting the lower bound of the amount of communication needed for all parties to obtain the value of a given function $f$ with successful probability $P_{succ}\geq 1-\delta$. We can naturally ask the inverse question: what is the highest successful probability for calculating the function $f$ if the amount of communication $C(f)$ is restricted to be upper bounded $C(f)\leq C_{bd}$ ? 
Note that unlike in the regular communication complexity problem where the bound of successful probability $1-\delta$ does matter so much, in this kind of CC problem, the communication bound is very important. Since there exists a trivial protocol for calculating arbitrary function $f$, for which Alice communicates her entire input to Bob, and thus $p_{succ}$ can always reach $1$ if allowed communication is greater than or equal to $\min\{|x|+1,|y|+1\}$.


There are two types of quantum communication complexity protocols: (i) preparation-measurement protocol and (ii) entanglement-assisted protocol; like the categorification of quantum key distribution protocol. In this work, we will mainly discuss the entanglement-assisted protocol.


The performance of a usual CC protocol $\mathcal{P}$ is characterized by the amount of communication, i.e., classical or quantum bits $C(\mathcal{P},f|p_{\text{succ}})$, required to achieve the success probability $p_{\text{succ}}$. The quantum advantage of CC problem means that there exists a quantum protocol $\mathcal{P}_Q$ such that for any classical protocol $\mathcal{P}_C$, we have $C(\mathcal{P}_C,f|p_{\text{succ}})>C(\mathcal{P}_Q,f|p_{\text{succ}})$.

The performance of the CC protocol $\mathcal{P}$ for calculating function $f$ can also be characterized by the maximal achievable success probability $p_{\text{succ}}(\mathcal{P},f|C_{bd})$ given a bounded amount of communication $C_{bd}$. Here, the communication could be classical bits or qubits, we say that there is a quantum advantage for ICC problem for calculating $f$ if there is a quantum protocol $\mathcal{P}_Q$ such that $p_{\text{succ}}(\mathcal{P}_Q,f|C_{bd})>p_{\text{succ}}(\mathcal{P}_C,f|C_{bd})$ for all classical protocol $\mathcal{P}_C$.

There is a simple and well-known example of CC problem proposed in Ref. \cite{buhrman2001quantum}, for which Alice and Bob receive bit strings $(x,a)\in \mathbb{B}^2$ and $(y,b)\in \mathbb{B}^2$ respectively and they
tend to calculate a function $f$ given by the language:
\begin{equation}\label{eq:CHSH}
L_{\text{Bell}}=\{(x,a;y,b)\in \mathbb{B}^2\times\mathbb{B}^2|a\oplus b=x\wedge y\}.
\end{equation}
All input strings distributed uniformly and two parties are allowed to exchange only two classical bits. Their goal is to calculate $L_{\text{Bell}}$ with as high successful probability as possible. In Ref. \cite{brukner2004bell}, Brukner \emph{et al.} present the optimal classical protocol and prove that using entangled quantum states that violate the CHSH inequality, the quantum solution of the problem has a higher success probability than the optimal classical protocol, thus exhibits the quantum advantage. The protocol works in the entanglement-assisted sense.

\section{From Bell inequality violation to quantum advantage for ICC problems}
\label{sec:qadv}

We now discuss how to translate a Bell test into an ICC problem using compatibility graph. To start with, let's consider the two-party case.
For a given experimental  compatibility graph $G^e$, which is a bipartite graph, and the vertices are labeled with $x_A=v_1,\cdots,v_n$ by Alice  and $x_B=u_1,\cdots, u_m$ by Bob respectively. There are some edges corresponding to $\gamma_{\langle ij\rangle}=1$ (called positive edges, drawn as black solid edge in Fig. \ref{fig:cycle})  and some others corresponding to $\gamma_{\langle ij\rangle}=-1$ (called negative edges, drawn as red dashed edge in Fig. \ref{fig:cycle}). 
We introduce a function which we refer to as target function
\begin{equation}
t(x_A,x_B)=\begin{cases}
0, \quad\text{for}\,\, \langle v_iu_j\rangle\,\, \text{positive edge},\\
1, \quad\text{for}\,\, \langle v_iu_j\rangle\,\, \text{negative edge}.
\end{cases}
\end{equation}
Consider the following two-party scenario, Alice and Bob receive $(x_A,y_A)$ and $(x_B,y_B)$ respectively, where $y_A,y_B =\pm 1$ and $x_A=1,\cdots,n$, $x_B=1,\cdots,m$, and the condition $\langle x_A x_B\rangle\in E(G^e)$, i.e. it is an edge of the experimental compatibility graph $G^e$, are promised. The function they are going to calculate is 
\begin{equation}
F_{G^e}(x_A,y_A;x_B,y_B)=y_Ay_B(-1)^{t(x_A,x_B)}.
\end{equation}
Notice that this is a partial function, for some inputs, the function is not defined, see Table \ref{tab:f3} for an example. In this way we can construct an CC function from arbitrary given experimental compatibility graph.

For the $n$-party case, the corresponding experimental graph is an $n$-partite hypergraph, the vertices of $k$-th party are labeled with $x_k=u^k_{1},\cdots,u^k_{m_k}$, the edge $\langle u^1_{i_1}\cdots u_{i_n}^n\rangle$ consists of $n$ vertices, one from each party. Similar as two-party case, we can define the target function
\begin{equation}
t(x_1,\cdots,x_n)=\begin{cases}
0, \quad\text{for}\,\, \langle u^1_{i_1}\cdots u^n_{i_n}\rangle\,\, \text{positive edge},\\
1, \quad\text{for}\,\,  \langle u^1_{i_1}\cdots u^n_{i_n}\rangle\,\, \text{negative edge}.
\end{cases}
\end{equation}
The function to be calculated is 
\begin{equation}
F_{G^e}(x_1,y_A;\cdots;x_n,y_n)=y_1\cdots y_n(-1)^{t(x_1,\cdots,x_n)}. \label{eq:ICCN}
\end{equation}

The CC problem to be solved is as follows, the $n$ parties try to calculate the function (\ref{eq:ICCN}), the $k$-th party receives the bit string $(x_k,y_k)$ with $x_k=u^k_1,\cdots,u^k_{m_k}$. The probability distribution for input strings is 
\begin{equation}
p(x_1,y_1;\cdots;x_n,y_n)=\frac{1}{2^n}\times \frac{1}{|E(G^e)|}.
\end{equation}
Each party is allowed to broadcast one classical bit of information, and $n$ parties broadcast the information simultaneously such that their broadcast bits are independent.

\begin{table}
	\caption{\label{tab:f3}The value of $F_{G^e_{6-cycle}}$, the columns are indexed by $(x_A,y_A)$ and the rows are indexed by $(x_B,y_B)$.}
	\begin{ruledtabular}
		\begin{tabular}{l|llllll}
			& (1,+1)  & (1, -1) & (2,+1) & (2,-1)  & (3,+1)  & (3,-1)\\
			\hline
			(1,+1)  & 1 & -1 & 1 &-1 & - & -\\
			(1, -1)  & -1 & 1 & -1& 1 & -& -\\
			(2,+1)  & - & -& 1 &-1 & 1&-1 \\
			(2,-1)   & - & -& -1& 1& -1& 1\\
			(3,+1)  & 1 &-1 &- & -& -1&1 \\
			(3,-1) & -1 & 1& -&- & 1& -1\\
		\end{tabular}
	\end{ruledtabular}
\end{table}

\subsection{Optimal classical protocol}

Let's now introduce an optimal classical protocol $\mathcal{P}_C$ for the above CC problem. To make things clearer, we take the two-party case as an example.
The main step is to calculate the target function part $(-1)^{t(x_A,x_B)}$. 
To do this, Alice and Bob firstly relabel their vertices as $x'_A$ and $x'_B$ such that the values $x'_A+x'_B$ are different for different edges. 
This can be done since $G^e$ is a finite graph. 
For example, for a fixed Bob's vertex $u_1$, the range of $u_1+v_i$ is $[N_1,N'_1]$, we can then set $u'_2>N'_1$, then all $u'_2+v_i>N'_1$, the intersection of ranges of $u_1+v_i$ and $u'_2+v_i$ is empty. 
By repeating the procedure $m$ times, we will achieve our goal. 
In fact, we can do more to relabel the vertices, such that the values corresponding to negative edges are odd numbers and the values corresponding to positive edges are even numbers. 
This is because that $v'_i+u'_j$ are now different for different edges. If the value is not as what we want, we can add a very large number to make the parity correct. 
In this way, we see that
\begin{equation}
(-1)^{t(x'_A,x'_B)}=(-1)^{x'_A+x'_B}.
\end{equation}
Before starting the calculation for a given experimental compatibility graph $G^e$, Alice and Bob firstly come together to discuss and fix the procedure to do the relabelling process. In fact, the easiest way is before calculation, we relabel the vertices as $x'_A$ and $x'_B$.

With the above preparation, we now present our classical protocol. 
Alice and Bob, when receiving inputs $(x_A,y_A)$ and $(x_B,y_B)$, choose to locally calculate two functions $a(x_A,\lambda_A)$ and $b(x_B,\lambda_B)$ such that $a(x_A,\lambda_A)=(-1)^{x'_A}$ and $a(x_A,\lambda_A)=(-1)^{x'_B}$. Note that here $\lambda_A,\lambda_B$ characterize their local classical resources and they may be classically correlated. Then Alice and Bob broadcast the results $e_A=y_Aa(x_A,\lambda_A)$ and $e_B=y_Bb(x_B,\lambda_B)$ respectively. After receiving the result, they both output with the answer function
\begin{equation}
\mathrm{Ans}_{\mathcal{P}_C}(x_A,y_A;x_B,y_B)=e_Ae_B.
\end{equation}
The success probability of the protocol is
\begin{align}
p_{\mathrm{succ}}(\mathcal{P}_C|C_{bd}=2)=&\frac{1}{|E(G^e)|}\big(\sum_{ \langle ij\rangle \,\,\text{positive}}p(ab=1|v_iu_j)\nonumber\\
&+ \sum_{ \langle ij\rangle\,\, \text{negative}}p(ab=-1|v_iu_j) \big). \label{eq:succp}
\end{align}
The protocol can achieve a success probability of $(B_C+|E(G^e)|)/2|E(G^e)|$, where $B_C$ is the classical bound for Bell inequality. For $2m$-cycle case, it's $p_{\mathrm{succ}}(\mathcal{P}_C|C_{bd}=2)=(2m-1)/2m$, especially for the well-known CHSH case $m=2$, $p_{\mathrm{succ}}(\mathcal{P}_C|C_{bd}=2)=3/4$.

For the $n$-party case, the protocol works similarly. The main difference is that the experimental compatibility graph is now an $n$-partite hypergraph. By relabeling the vertices, we have
\begin{equation}
(-1)^{t(x_1,\cdots,x_n)}=(-1)^{x'_1+\cdots +x'_n}.
\end{equation}
After receiving the input bit strings, each party chooses to locally calculate a function $e_k=y_i a(x_k,\lambda_k)$ with $a_k(x_k,\lambda_k)=(-1)^{x'_k}$. Finally they broadcast $e_k$ simultaneously and output the value
\begin{equation}
\mathrm{Ans}_{\mathcal{P}_C}(x_1,y_1;\cdots ;x_n,y_n)= e_1\cdots e_n.
\end{equation} 
The success probability is similar to Eq. (\ref{eq:succp}). The protocol can achieve a success probability of $(B_C+|E(G^e)|)/2|E(G^e)|$, where $B_C$ is the classical bound for Bell inequality. In this protocol each party indeed only broadcasts one classical bit of information.

Before we talk about the quantum advantage of the entanglement-assisted protocol, we need to prove that this is in fact the optimal classical protocol.

\emph{Proof of the optimality of the protocol.}\textemdash We now show that the above protocol is optimal, i.e., there is no classical protocol reaching a higher success probability.
For the two-party case, what we need to show is that, when Alice and Bob initially share classical randomness, there is no $C_{bd}=2$ protocol for which Alice and Bob can calculate the function $F_{G^e}$ with success probability greater than $(B_C+|E(G^e)|)/2|E(G^e)|$. 
Firstly, we observe that an $n$-bit Boolean function $f(x_1,\cdots,x_n)$ with values $\pm 1$ can be decomposed as 
\begin{equation}
f(x_1,\cdots,x_n)=\sum_{i_1,\cdots,i_n=0,1}T_{i_1,\cdots,i_n}x_1^{i_1}\cdots x_m^{i_n}. \label{eq:exp}
\end{equation}
Since $f(x_1,\cdots,x_n)=\pm 1$, we have $|T_{i_1,\cdots,i_n}|\leq 1$. In fact, the expansion coefficients are given by
\begin{equation}
T_{i_1,\cdots,i_n}=\frac{1}{2^n}\sum_{x_1,\cdots,x_n=\pm 1}f(x_1,\cdots,x_n)x_1^{i_1}\cdots x_n^{i_n}.
\end{equation}
Now consider the function $F_{G^e}(x_A,y_A;x_B,y_B)$, for convenience, we introduce the new variables $\tilde{x}_A=(-1)^{x'_A}$ and  $\tilde{x}_B=(-1)^{x'_B}$. Using the expansion of Eq.(\ref{eq:exp}), the broadcast bits become 
\begin{equation}
e_{i}=e_i(\tilde{x}_i,y_i)= (T_{00}+T_{10}\tilde{x}_i) + (T_{01} +T_{11}\tilde{x}_i)y_i=
c_i(\tilde{x}_i)+d_i(\tilde{x}_i)y_i,
\end{equation}
where $|c_i(\tilde{x}_i)|+|d_i(\tilde{x}_i)|=1$ and $|c_i(\tilde{x}_i)|,|d_i(\tilde{x}_i)|=0,1$, with $i=A,B$. 
The inner product of the Alice's answer function with function $F_{G^e}$ can be defined as 
\begin{align}
\langle \mathrm{Ans}_A,F_{G^e}\rangle 
=\sum_{\scriptsize{\begin{array}{ll}x_A,y_A,\\ x_B,y_B\end{array}}}
\frac{\mu({x}_A,{x}_{B})}{4}\mathrm{Ans}_A({x}_A,y_A,e_B)F_{G^e}({x}_A,y_A;{x}_B,y_B). 
\end{align}
Here $\mu({x}_A,{x}_B)/4$ is the probability distribution over inputs.
We see that when $\mathrm{Ans}_A({x}_A,y_A,e_B)=F_{G^e}({x}_A,y_A;{x}_B,y_B)$, they contribute $+1$ in the above summation, otherwise they contribute $-1$. Notice the fact that $1=\sum_{x_A,y_A, x_B,y_B}\frac{\mu({x}_A,{x}_{B})}{4}$, the success probability for Alice to output the correct answer can thus be written as $p_{\mathrm{succ}}=\frac{1}{2}(1+\langle \mathrm{Ans}_A,F_{G^e}\rangle)$.
Inserting the expression of $F_{G^e}$ and the expansion $\mathrm{Ans}_A({x}_A,y_A,e_B)=\mathrm{Ans}_A(\tilde{x}_A,y_A,e_B)=\sum_{j_{\tilde{x}}j_yj_e}T_{j_{\tilde{x}}j_yj_e}\tilde{x}_A^{j_{\tilde{x}}}y_A^{j_y}e_B^{j_e}$ into it, we obtain
\begin{equation}
p_{\mathrm{succ}}=\frac{\sum_{{x}_A, {x}_B}(-1)^{t(x_A,x_B)}(T_{011}+T_{111}\tilde{x}_A)d_B(\tilde{x}_B)}{|E(G^e)|}.
\end{equation}
From the definition of the expansion coefficients we have $|T_{011}+T_{111}\tilde{x}_A|\leq 1$. Using the Bell inequality, for arbitrary functions  $f(x_A)$, $g(x_B)$ with  $|f(x_A)|, |g(x_B)|\leq 1$ we have
\begin{equation}
\sum_{x_A,x_B} \gamma_{\langle x_A x_B \rangle} f(x_A) g(x_B)= \sum_{x_A,x_B}(-1)^{t(x_A,x_B)}     f(x_A) g(x_B)  \leq \frac{B_C+|E(G)|}{2}.
\end{equation}
Thus the success probabity must satisfy $p_{\mathrm{succ}}\overset{C}{\leq} \frac{B_C+|E(G^e)|}{2|E(G^e)|}$. Since the protocol we gave before reaches the bound, it is the optimal classical protocol.
Similarly for Bob, we can define $\langle \mathrm{Ans}_B,F_{G^e}\rangle$. From symmetry of the problem expression, the same result holds for Bob.
For the $n$-party case, the proof is completely the same.

The proof here is in the same spirit with the proof in the Ref. \cite{brukner2004bell}. Another way to prove the optimality is using the traditional communication complexity theoretic approach, for which we first prove a lower bound of the deterministic protocol. Then using a famous theorem \cite{kushilevitz_nisan_1996} which states that the communication complexity $R_{\epsilon}(f)$ of the randomized protocol for computing a function $f$ with error $\epsilon$ has a relationship with the communication complexity $D_{\epsilon}(f|\mu)$ of deterministic protocol for computing the function $f$ with error $\epsilon$ for which inputs distributed with $\mu$ as: $R_{\epsilon}(f)=\max_{\mu} D_{\epsilon}(f|\mu)$, the lower bound of the deterministic protocol can be proved by assuming a protocol-tree with depth 2 (for two-party case) and discussing the partitions of the inputs by different nodes of the protocol-tree.

\subsection{Entanglement-assisted protocol}
The quantum protocol works as follows. We take two-party case as an example. Alice and Bob preshare an entangled quantum state $|\psi\rangle_{AB}$, upon which Alice and Bob can choose $\pm 1$-valued observables $A_1,\cdots,A_m$ and $B_1,\cdots,B_n$ and obtain a violated value of Bell inequality corresponding the the experimental  compatibility graph $G^e$. Now if Alice and Bob receive input values $x_A=v_i$ and $x_B=u_j$, they can measure the corresponding observables $A_i$ and $B_j$ and output $a_A=a_i$ and $b_B=b_j$. Then Alice and Bob broadcast the classical bits $e_A=y_Aa_A$ and $e_B=y_Bb_B$ respectively. After receiving the communicated bits, Alice and Bob both give their answers as $\mathrm{Ans}_A=\mathrm{Ans}_B=e_Ae_B$.
The success probability is still Eq. (\ref{eq:succp}). We see that it can exceed the bound of classical protocol, thus exhibits the quantum advantage.

To make it more clear, let us first take $G_{2m-cycle}^e$ as an example (see Example \ref{exp:1}). Suppose that Alice and Bob preshare the singlet state $|\psi^-\rangle=\frac{1}{\sqrt{2}}(|01\rangle-|10\rangle).$
The observables for Alice are  $A_i=\mathbf{m}_i\cdot \bm{\sigma}$, where
\begin{equation}
\mathbf{m}_i=(\cos\frac{ (2i-1)\pi}{2m},0,\sin \frac{ (2i-1)\pi}{2m}), \,\, i=1,\cdots, m,\nonumber
\end{equation}
and for Bob are  $B_j=\mathbf{n}_j\cdot \bm{\sigma}$, where
\begin{equation}
\mathbf{n}_j=(\cos\frac{ j\pi}{m},0,\sin \frac{ j\pi}{m}), \,\, j=1,\cdots, m.\nonumber
\end{equation}
With these measurements, Alice and Bob can achieve a success probability $p_{\mathrm{succ}}=\frac{\cos \pi/2m+1}{2}$, which corresponds to Tsirelson bound of the $2m$-cycle Bell inequality. We see that, the success probability is a monotone increasing function, and when $m\to \infty$, it tends to $1$.

Another example is the $G^e_{2|3}$ as illustrated in Example \ref{exp:2}. Alice and Bob still preshare the singlet state, and Alice chooses to measure $A_{1}=\sigma_x$ and $A_2=\sigma_z$, Bob chooses to measure $B_{j}=\mathbf{n}_j \cdot \bm{\sigma}$ with
\begin{equation}
\begin{aligned}
&\mathbf{n}_1=(\cos \frac{\pi}{4}, 0, \cos \frac{\pi}{4}), \\
&\mathbf{n}_1=(\cos \frac{3\pi}{4}, 0, \cos \frac{3\pi}{4}), \\
&\mathbf{n}_3=(\cos (\frac{\pi}{4}+\theta), 0, \cos (\frac{3\pi}{4}+\theta)),  \theta \ll 1.
\end{aligned} \nonumber
\end{equation}
The optimal classical protocol can achieve a success probability of $5/6$. Here the quantum protocol can almost reach the success probability of $(3\sqrt{2}+6)/12$ for $\theta$ small enough, which is greater than the success probability for optimal classical protocol, thus it exhibits quantum advantage.
Notice that the above problem is closely related to the problem of simulation of nonlocal correlation via classical communication \cite{Toner2003communication}. Our result here matches well with former result that by two bits of classical communication, the Bell nonlocal measurement statistics can be simulated.

\subsection{Popescu-Rohrlich box protocol}
Let us now consider a non-signaling world which is beyond quantum mechanics. Suppose that Alice and Bob preshare a black box such that for the positive edge $\langle ij\rangle$ of experimental  compatibility graph $G^e$, the probability distribution of outputs for measurements $A_i,B_j$ is
\begin{equation}
p(a_i,b_j|A_i,B_j)=\begin{cases}
1/2, \quad a_ib_j=1,\\
0,\quad a_ib_j=-1.
\end{cases}
\end{equation}
And for negative edges, the distribution is
\begin{equation}
p(a_i,b_j|A_i,B_j)=\begin{cases}
0, \quad a_ib_j=1,\\
1/2,\quad a_ib_j=-1.
\end{cases}
\end{equation}
This kind of black box is known as Popescu-Rohrlich box \cite{Popescu1994} or perfect nonlocal box. It's easily checked that the box satisfies the non-signaling principle.

With the help of Popescu-Rohrlich box, we can reach a success probability $p_{\mathrm{succ}}=1$. The protocol works similarly as the entanglement-assisted protocol. After receiving the inputs $x'_A=v_i$ and $x'_B=u_j$, Alice and Bob choose to measure $A_i$ and $B_j$ jointly and output $a_i$ and $b_j$ with probability $p(a_i,b_j|A_i,B_j)$. After many runs of the experiment, Alice and Bob check their success probability, it's obvious from Eq.(\ref{eq:succp}) that for the Popescu-Rohrlich box, the success probability is $p_{\mathrm{succ}}=1$. This matches well with the result in Refs. \cite{van2005implausible,Brassard2006limit} which states that using perfect nonlocal box can make CCP trivial for arbitrary Boolean function.

\section{Conclusions and discussions}
To find the bound of classical theory and quantum theory is of great importance for understanding the nature of our universe. In this work, we try to understand the problem from a communication-complexity theoretic perspective. 
By restricting the classical communications, two parties can calculate a given function with different success probabilities, this shows that the strength of quantum correlations is much stronger than the classical one. 
These results shed new light on the bound between classical and quantum worlds.
From a practical point of view, our result provides a method to construct CC function from an arbitrary given experimental compatibility graph or hypergraph. When the graph is bipartite graph, it gives a two-party CC function, when the graph is multipartite, it gives a multi-party CC function. Our construction may have potential applications in the practical CC problems where one wants to extract quantum advantages from Bell nonlocality.

\begin{acknowledgements}
Z. A. Jia acknowledges Zhenghan Wang and the math department of UCSB for hospitality during his visiting at UCSB where some parts of work are carried out.
\end{acknowledgements}

\bibliographystyle{apsrev4-1-title}
%

\end{document}